\documentclass[aps,pre,onecolumn,superscriptaddress,groupedaddress,showpacs]{revtex4}
\usepackage{graphicx,epsf}
\begin{document}
%----------------------------------------------------------------------------
%                              TITLE
%----------------------------------------------------------------------------
\title{Shape anisotropy of polymers in disordered environment}
%----------------------------------------------------------------------------
%                              AUTHORS revtex4-style
%----------------------------------------------------------------------------
\author{Viktoria Blavatska}
\email[]{E-mail:  Viktoria.Blavatska@itp.uni-leipzig.de; viktoria@icmp.lviv.ua}
\affiliation{Institut f\"ur Theoretische Physik and Centre for Theoretical Sciences (NTZ),\\ Universit\"at Leipzig, Postfach 100920,
D-04009 Leipzig, Germany}
\affiliation{Institute for Condensed
Matter Physics of the National Academy of Sciences of Ukraine,\\
79011 Lviv, Ukraine}
\author{Wolfhard Janke}
\email[]{E-mail: Wolfhard.Janke@itp.uni-leipzig.de}
\affiliation{Institut f\"ur Theoretische Physik and Centre for Theoretical Sciences (NTZ),\\ Universit\"at Leipzig, Postfach 100920,
D-04009 Leipzig, Germany}
%----------------------------------------------------------------------------
%                             ABSTRACT
%----------------------------------------------------------------------------

\begin{abstract}

We study the influence of structural obstacles in a disordered environment on the size and shape characteristics of
long flexible polymer macromolecules. We use the model of self-avoiding random walks on
diluted regular lattices at the percolation threshold in space dimensions $d=2$, $3$. Applying the
Pruned-Enriched Rosenbluth Method (PERM),
we numerically estimate rotationally invariant universal quantities such as the averaged asphericity $\langle A_d \rangle$ and prolateness $\langle S \rangle$
of polymer chain configurations. Our results quantitatively reveal the extent of anisotropy of macromolecules due to the presence of
structural defects.
\end{abstract}
% 36.20.-r:  Macromolecules and polymer molecules
%67.80.dj    Defects, impurities, and diffusion
% 64.60.ah percolation in phase transitions
% C07.05.Tp omputer modeling and simulation

%\date{\today}
\maketitle%

\section{Introduction}

Topological properties of macromolecules, such as the shape and size of a typical polymer chain configuration, are of interest in various respects.
 The shape of proteins affects their folding dynamics and motion
in a cell and is relevant in comprehending complex cellular phenomena, such as
catalytic activity \cite{Plaxco98}. The hydrodynamics of polymer fluids is essentially affected by the size and shape of individual
macromolecules \cite{Neurath41}; the polymer shape plays an important role in determining its molecular weight
in gel filtration chromatography \cite {Erickson09}.

%In particular, it was shown \cite{Hanley83} that averaged shape of polymer configuration is an important parameter in a
%phenomenogical theory describing experiments on polymeric fluid flow at high strain rates.
Already in 1934 it was realized \cite{Kuhn34} that viscous properties of polymer solutions are significantly
different than predicted by the theory for dissolved sphere-like molecules:
flexible polymer chains in good solvents form crumpled coil shapes, which are surprisingly anisotropic.
Since then, a considerable amount of work has been done in exploring size and shape characteristics
of various macromolecules \cite{Domb69,Witten78,Solc71,Kranbuehl77,Benhamou85,Aronovitz86,
Rudnick86,Bishop88,Honeycutt88,Cardy89,Cannon91,Jagodzinski92,Dima04,Rawat09}.

Solc and Stockmayer \cite{Solc71} introduced as a shape measure of macromolecules the normalized average eigenvalues $\lambda_i$ of the
gyration tensor.
Numerical simulations in $d=3$ dimensions give $\{\langle \lambda_1\rangle$,
 $\langle \lambda_2\rangle$, $ \langle \lambda_3\rangle \}$=$\{0.790,0.161,0.054 \}$ \cite {Kranbuehl77}, indicating a high anisotropy
 of typical polymer
configurations compared with the purely isotropic case $\{1/3$, $1/3$, $1/3 \}$.
To compute the quantities $\lambda_i$ analytically is, however,
difficult, because one must explicitly diagonalize the gyration tensor for each realization in an ensemble of polymers.
It was proposed \cite{Aronovitz86,Rudnick86} to characterize the asymmetry of polymer configurations by rotationally invariant universal
 quantities,
such as the averaged asphericity $\langle A_d \rangle$ and prolateness $\langle S \rangle$.
$\langle{A_d}\rangle$ takes on a maximum value of one for a completely stretched, rod-like configuration,
and equals zero for spherical form, thus obeying the inequality: $0\leq \langle A_d \rangle \leq 1$. The quantity $\langle S \rangle$, defined in $d=3$,
 takes on a positive value for prolate ellipsoid-like configurations,  and is negative for oblate shapes, being
 bounded to the interval $-1/4 < \langle S \rangle < 2$.
% (detailed definitions of both $\langle A_d \rangle $ and $\langle S \rangle$ will be given below).
To characterize the size measure of a single flexible polymer chain, one usually considers the mean-squared end-to-end distance $\langle R_e^2 \rangle $
and radius of gyration
$\langle R_G^2 \rangle$, both
governed by the same scaling law: $ \langle R_e^2 \rangle \sim \langle R_G^2 \rangle \sim  N^{2\nu},
$
where  $N$ is the mass of the macromolecule (number of monomers in a polymer chain) and $\nu$ a universal exponent ($\nu>1/2$ $(d<4)$, $\nu=1/2$ $(d\geq4)$).
The ratio of these two characteristic distances, the so-called size ratio $g_d\equiv\langle R_e^2 \rangle/\langle R_G^2 \rangle$, also appears to be a universal,
rotationally invariant quantity ($g_d>6$ $(d<4)$, $g_d=6$ $(d\geq 4))$ \cite{Witten78}.

Numerous studies indicate that a  typical flexible polymer chain in good solvent takes on the shape of an elongated, prolate ellipsoid. In particular,
using the data of x-ray crystallography and cryo-electron microscopy, it was found that the majority of nonglobular proteins are characterized by $A_3$ values from $0.5$ to $0.7$
and $S$ values from $0$ to $0.6$ \cite{Dima04,Rawat09}. The shape parameters of polymers were analyzed analytically, based on the renormalization group approach \cite{Benhamou85,Aronovitz86,Jagodzinski92}, and  estimated in numerical simulations \cite{Domb69,Solc71,Bishop88,Honeycutt88}.
Previous estimates of the shape and size characteristics of flexible polymer chains in $d=2$, $3$ are given in Table \ref{dani}.

\begin{table}[b!]
\caption{ \label{dani} Size ratio, averaged asphericity  and prolateness of flexible polymer chains on regular lattices.
 $a$: Ref. \cite{Domb69}, $b$:  Ref. \cite{Bishop88}, $c$:  Ref. \cite{Jagodzinski92}. }
%\begin{center}
\begin{tabular}{cccc}
\hline
 $d$ & $g_d$ & $\langle A_d \rangle$& $\langle S \rangle$\\
\hline
2 &  $7.14\pm0.03^a$ & $0.501\pm 0.003^b$  & -- \\
3 & $6.249\pm0.03^c$ & $0.431\pm0.002^c$ & $0.541\pm0.004^c$ \\
\hline
\end{tabular}
%\end{center}
\end{table}

In real physical processes, one is often interested in the question as to how structural obstacles (impurities) in the environment
alter the behavior of a system. The density fluctuations of obstacles lead to a large spatial inhomogeneity and create pore spaces,
which are often of fractal structure \cite{Dullen79}.
In polymer physics, of great importance is understanding of the behavior of macromolecules in the presence of structural disorder,
 e.g., in colloidal solutions \cite{Pusey86} or microporous membranes \cite{Cannel80}.
In particular, a related problem is relevant when studying the protein folding dynamics in the cellular environment \cite{Kumarrev}.
Biological cells can be described as a highly disordered environment due to the presence of a large amount of soluble and insoluble biochemical species, which
occupy up to $40\%$ of the total aquabased volume \cite{Minton01}.
Structural obstacles strongly effect the protein folding and aggregation \cite{Horwich,Winzor06,Kumar,Echeverria10}.
Recently, it was realized experimentally \cite{Samiotakis09} that macromolecular crowding has a dramatic effect
on the shape properties of proteins.
% vstavka
To explain the physics behind the macromolecular crowding effects, a statistical theory of
excluded volume interaction between proteins and hard particles was developed \cite{Minton05}.~The  folding of proteins in
crowded environment was studied numerically by off-lattice
polypeptide chain simulations in the presence of repulsive spherical particles in Refs.
 \cite{Homouz08,Cheung05}.
% kinec vstavky

In the language of lattice models, the disordered environment with structural obstacles can be considered as a lattice where some amount
of randomly chosen sites contain defects, which are to be avoided by the polymer chain. Of particular interest is the case, when
the concentration  of lattice sites allowed for the polymer chain
equals the critical concentration $p_c=0.592746$ $(d=2)$ \cite{Ziff94}, $p_c=0.31160$ $(d=3)$ \cite{Grassberger92} and the
lattice becomes percolative: a percolation cluster, having a fractal structure,
occurs \cite{Stauffer}.
% vstavka
The lattice model with percolation has played a key role in statistical physics for describing
structurally disordered systems. This approach is limited, however, in the realm of
biological systems and processes. The core of this difficulty is the highly specialized and nonequilibrium nature
of the biological world. Still, some aspects such as the  universal configurational properties of bioproteins
 seem to be amenable to analyses via this simplified model \cite{Kumarrev,Kumar}.
% kinec vstavky

Studying processes on percolative lattices, one encounters two possibilities.
In the  first, one considers  only percolation clusters with linear size much larger than the typical length of the physical phenomenon under discussion (polymer chain length in
our case). The other statistical ensemble includes all lattice sites free of defects, which can be found in a percolative lattice.
In the latter case, the polymer can be trapped (localized) in confined regions of pure sites, and so-called localization phenomena occur which lead to
{\it decreasing} the size of the macromolecule. Such a situation  has been studied in Ref. \cite{Honeycutt88}, realizing the
shrinking of polymer shapes with a trend to {\it decreasing} anisotropy.
In what follows, we will be interested in the former case, when a polymer chain resides only on the percolation cluster of fractal structure;
this results in {\it increasing} the swelling of the polymer coil compared with the pure solution case \cite{Woo91,Grassberger93,Rintoul94,Ordemann02,Janssen07,Blavatska08}.
However, the important question of how do the shape parameters of a polymer chain change quantitatively, when the polymer is located on a fractal cluster,
still is completely unresolved.

The purpose  of the present paper is to report computer simulations of a model of flexible polymer chains on the backbone of fractal percolation clusters
in $d=2$, $3$. We aim to obtain  numerical estimates of $\langle A_d^{p_c} \rangle$, $\langle S^{p_c} \rangle$ and $g^{p_c}_d$,
and thus quantitatively describe the change
in asymmetry of typical polymer configurations due to the presence of structural obstacles.
The rest of the paper is organized as follows: in the next section we introduce the discretized mathematical model of polymer chain and Section
III describes the details of our computer simulations. We give discussions of our results in Section IV and end up by giving conclusions and an outlook.

\section{Description of the model}

Let $\vec{R}_n=\{x_n^{1},\ldots,x_n^d\}$ be the position vector of the $n$th monomer of a polymer chain ($n=1,\ldots,N$).
The measure of the shape properties of a specified spatial conformation of the chain can be characterized \cite{Solc71,Aronovitz86,Rudnick86}
in terms of the gyration tensor $\bf{Q}$ with components:
\begin{equation}
Q_{ij}=\frac{1}{N}\sum_{n=1}^N(x_n^i-{x^i_{CM}})(x_n^j-{x^j_{CM}}),\,\,\,\,\,\,i,j=1,\ldots,d,
\label{mom}
\end{equation}
with ${x^i_{CM}}=\sum_{n=1}^Nx_n^i/N$ being the coordinates of the center-of-mass position vector ${\vec{R}_{CM}}$.

The spread in eigenvalues $\lambda_i$ of the gyration tensor describes the distribution of monomers inside the polymer coil and
thus measures the asymmetry of a molecule; in particular, for a symmetric (spherical)
configuration all the eigenvalues $\lambda_{i}$ are equal.

It was found convenient to characterize the shape of polymers by rotationally invariant
universal combinations of components of the gyration tensor \cite{Aronovitz86,Rudnick86}.
The first invariant of $\bf{Q}$ is the squared radius of gyration
\begin{equation}
R_G^2 =\frac{1}{N} \sum_{n=1}^N (\vec{R}_n-{\vec{R}_{CM}})^2  =  \sum_{i=1}^d Q_{ii} = {\rm Tr}\, \bf{Q}, \label{rg}
\end{equation}
which measures the average distribution of monomers with respect to the center of mass.
Let ${\overline{\lambda}}\equiv {\rm Tr}\, {\bf{Q}}/d$
be the average eigenvalue of  the gyration tensor.
Then the extent of asphericity of a polymer chain configuration is characterized by the quantity ${A_d}$ defined as \cite{Aronovitz86}:
\begin{equation}
{A_d} =\frac{1}{d(d-1)} \sum_{i=1}^d\frac{(\lambda_{i}-{\overline{\lambda}})^2}{\overline{\lambda}^2}=
\frac{d}{d-1}\frac{\rm {Tr}\,{\bf{\hat{Q}}}^2}{(\rm{Tr}\,{\bf{Q}})^2}, \label{add}
\end{equation}
with ${\bf{{\hat{Q}}}}\equiv{\bf{Q}}-\overline{\lambda}\,{\bf{I}}$ (here $\bf{I}$ is  the unity matrix).
This universal quantity equals zero for a spherical configuration, where all the
 eigenvalues are equal, $\lambda_i=\overline{\lambda}$, and takes a maximum value
of one in the case of a rod-like configuration, where all the eigenvalues equal zero except of one.
Thus, the inequality holds: $0\leq A_d\leq 1$.
Another rotationally invariant quantity, defined in three dimensions, is the so-called prolateness $S$ \cite{Aronovitz86,Rudnick86}:
\begin{eqnarray}
S =\frac{\prod_{i=1}^3(\lambda_{i}-{\overline{\lambda}})}{{\overline{\lambda}}^3}=27\frac{{\rm det}\, \bf{{\hat{Q}}}}{(\rm{Tr}\,{\bf{Q}})^3}. \label{sdd}
\end{eqnarray}
If the polymer is absolutely prolate, rod-like  ($\lambda_1\neq0,\lambda_2=\lambda_3=0$), it is easy to see that $S$ equals two.
For absolutely oblate, disk-like conformations  ($\lambda_1=\lambda_2,\lambda_3=0$), this quantity takes on a value of $-1/4$.
In general, $S$ is positive for prolate ellipsoid-like polymer conformations  ($\lambda_1\gg \lambda_2\approx\lambda_3$) and negative for oblate
ones ($\lambda_1\approx\lambda_2\gg\lambda_3$), whereas its magnitude measures how oblate or prolate the polymer is.
Note that since $\overline{\lambda}$ and the quantities in (\ref{rg})-(\ref{sdd}) are expressed in terms of rotationally invariants, there is no need to explicitly determine the
eigenvalues $\lambda_i$ which simplifies the numerics significantly.

The average of quantities (\ref{rg})-(\ref{sdd}) for a given polymer chain length $N$, denoted as $\langle \ldots \rangle$, is performed over an
ensemble of possible configurations of a chain.
 Note that some analytical and numerical treatments avoid the averaging of the  ratio in (\ref{add}), (\ref{sdd})
and evaluate quantities:
\begin{equation}
\hat{A}_d=\frac{1}{d(d-1)}  \sum_{i=1}^d\frac{\left\langle(\lambda_{i}-{\overline{\lambda}})^2\right\rangle}{\left\langle\overline{\lambda}^2\right\rangle},\,\,\,\,\,\,\,
\hat{S}= \frac{\left\langle\prod_{i=1}^3(\lambda_{i}-{\overline{\lambda}}) \right\rangle}{\left\langle{\overline{\lambda}}^3\right\rangle}, \label{avratio}
\end{equation}
which should be distinguished from the averaged asphericity and prolateness:
\begin{equation}
\langle{A_d}\rangle =\frac{1}{d(d-1)} \left \langle \sum_{i=1}^d\frac{(\lambda_{i}-{\overline{\lambda}})^2}{\overline{\lambda}^2} \right\rangle,\,\,\,\,\,\,\,
\langle{S}\rangle=\left\langle \frac{\prod_{i=1}^3(\lambda_{i}-{\overline{\lambda}})}{{\overline{\lambda}}^3} \right\rangle. \label{avad}
\end{equation}
Contrary to $\langle{A_d}\rangle$ and $\langle{S}\rangle$, the quantities (\ref{avratio}) have no direct
relation to the probability distribution of the shape parameters $A_d$ and $S$.  As pointed out by Cannon {\em et al.} \cite{Cannon91}, this definition
 overestimates the influence of larger polymer configurations on the mean shape properties and suppresses the influence of compact ones.
This artificially leads to overestimated values for shape parameters. The difference between $\langle{A_d}\rangle$ and $\hat{A}_d$  on
regular lattices was
found to be really large
($\hat{A}_2=0.625\pm0.008$, $\hat{A}_3=0.546\pm0.008$ \cite{Bishop88}, which should be
compared with the data in Table \ref{dani}).

\section{The method}

We start with regular lattices with sites assigned to contain an obstacle with probability $1-p_c$ and be
allowed for the polymer chain otherwise.
To obtain the backbone of a percolation cluster on a given disordered lattice, we apply an algorithm consisting of the following two steps: first finding the percolation cluster
based on the  site-labeling method of Hoshen and Kopelman \cite{Hoshen76} and then
 extracting the backbone of this cluster \cite{Porto97} (the algorithm is explained in detail in our previous papers
\cite{Blavatska08}). Since the self-avoiding walk trajectory can be trapped in dangling ends of percolation clusters,
the infinitely long chains can only exist on the backbone of the cluster.
The question about differences in universal configurational
properties of SAWs walking on a percolation cluster and its backbone
was addressed recently, e.g., in Ref. \cite{Blavatska08}. We constructed 1000 clusters in each space dimension.

To study shape properties of typical polymer chain configurations modelled by self-avoiding random walks (SAWs) on the constructed percolation clusters, we use the pruned-enriched Rosenbluth
method (PERM) \cite{Grassberger97}, combining the original Rosenbluth-Rosenbluth (RR) algorithm of growing chains \cite{Rosenbluth55}
and population control \cite{Wall59}. The growth process starts at the center of the percolation cluster, and each $n$th monomer is placed at
a randomly chosen neighbor site of the last placed $(n-1)$th  monomer ($n\leq N$, where $N$ is total length of the polymer).
If this randomly chosen site is already visited by a chain trajectory or does not belong to the percolation cluster,
it is avoided without discarding the chain, but the bias is corrected by means of giving a weight $W_n\sim (\prod_{l=1}^n m_l)$
to each sample configuration at the $n$th step, where $m_l$ is the number of free lattice sites to place the $l$th monomer.
The growth is stopped when the total length $N$ of the chain is reached, then the next chain is started to grow from the same starting point.
The configurational averaging for any quantity of interest  then has the form:
\begin{eqnarray}
&&\langle (\ldots) \rangle=\frac{1}{Z_N}{\sum_{{\rm conf}}W_N^{{\rm conf}}(\ldots)},
\,\,\,\,Z_N=\sum_{{\rm conf}} W_N^{{\rm conf}} \label {R}.
\end{eqnarray}

The Rosenbluth method, however, also suffers from attrition: if all next neighbors at some step ($n<N$) are occupied, i.e., the chain is running
into a ``dead end", the complete chain has to be discarded and the growth process has to be restarted. Grassberger \cite{Grassberger97} proposed a considerable
improvement of the efficiency by increasing the number of successfully generated chains. The weight fluctuations of the growing chain are suppressed
in PERM by pruning configurations with too small weights, and by enriching the sample with copies of high-weight configurations. These copies are made
while the chain is growing, and continue to grow independently of each other. Pruning and enrichment are performed by choosing thresholds $W_n^{<}$
and $W_n^{>}$ depending on the estimate of the partition sum for the $n$-monomer chain. These thresholds are
continuously updated as the simulation progresses. If the current weight $W_n$ of an $n$-monomer chain is less than $W_n^{<}$,
 the chain is discarded with probability $1/2$, otherwise it is kept and its weight is doubled.  If $W_n$ exceeds
$W_n^{>}$, the configuration is doubled and the weight of each identical copy is taken as half the original weight.  For a value of the weight
 lying between the thresholds, the chain is simply continued without enriching or pruning the sample.
For updating the threshold values we apply similar rules as in \cite{Hsu03,Bachmann03}: $W_n^{>}=C(Z_n/Z_1)(c_n/c_1)^2$ and $W_n^{<}=0.2W_n^{>}$, where $c_n$ denotes the number of created chains having length $n$, and the parameter $C$ controls the pruning-enrichment statistics.
After a certain number of chains of total length $N$ is produced, the given tour  is finished and a new one starts. The pruning-enrichment
 control is adjusted  such that on average 10 chains of total length $N$ are generated per each tour \cite{Bachmann03}.

For estimations of quantities of interest we have to perform two types of averaging:  the first over all polymer configurations
on a single percolation cluster according to  (\ref{R});
the second average is carried out over different realizations of disorder, i.e., over all percolation clusters constructed:
\begin{eqnarray}
&&\overline{\langle \ldots \rangle}{=}\frac{1}{M}\sum_{i{=}1}^M \langle \ldots \rangle_i,\label{av}
\end{eqnarray}
where $M$ is the number of different clusters and the subscript $i$ means that a given quantity is calculated on the cluster $i$.

\begin{figure}[b!]
\includegraphics[width=6cm]{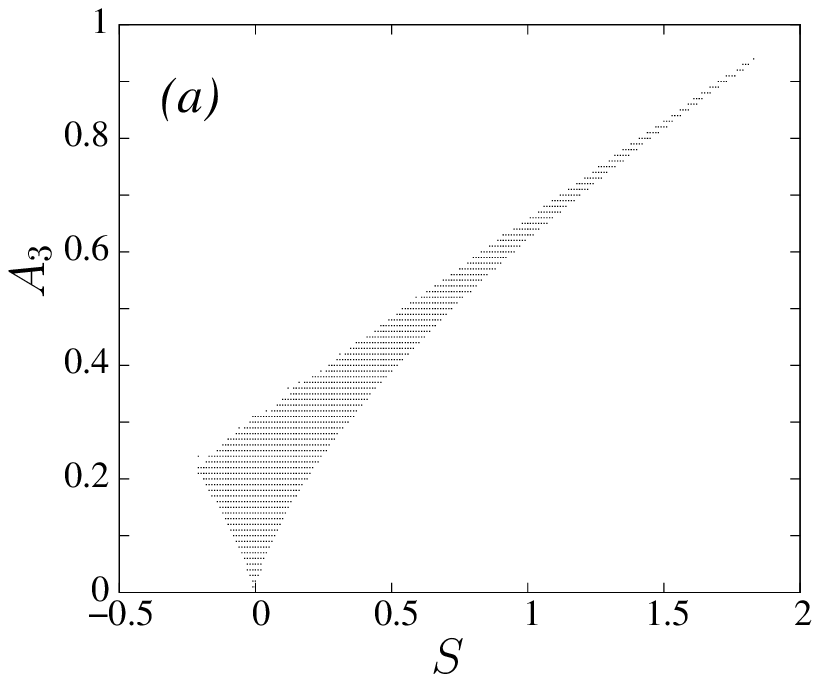}
\hspace*{0.5cm}
\includegraphics[width=6.3cm,bb=197 604 429 789]{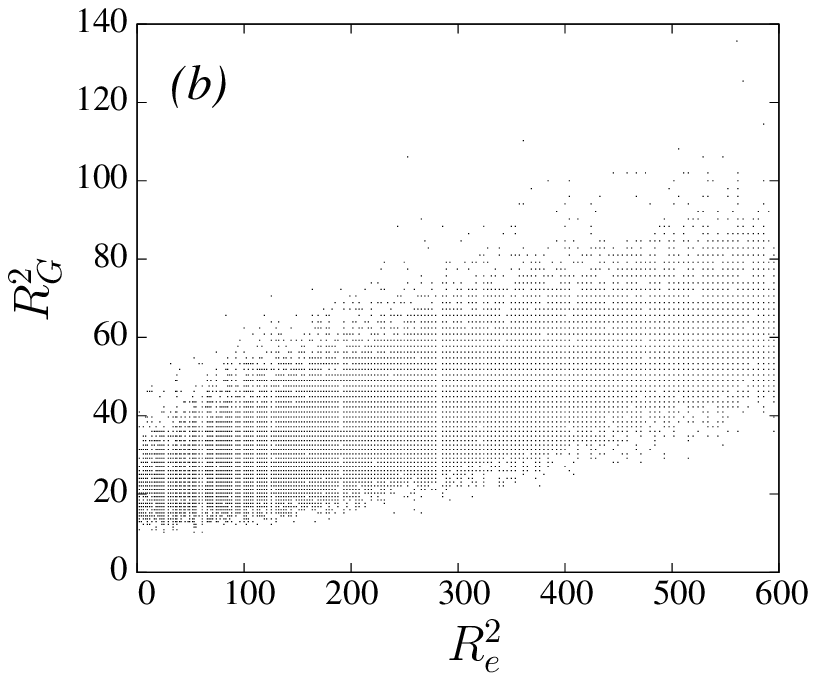}
\caption{ \label{correlation} Correlation between $A_3$ and $S$ values (a) and between $R_G^2$ and $R_e^2$
 values (b) in $d=3$ for chain length $N=120$. }
\end{figure}

\begin{figure}[t!]
\includegraphics[width=5.8cm]{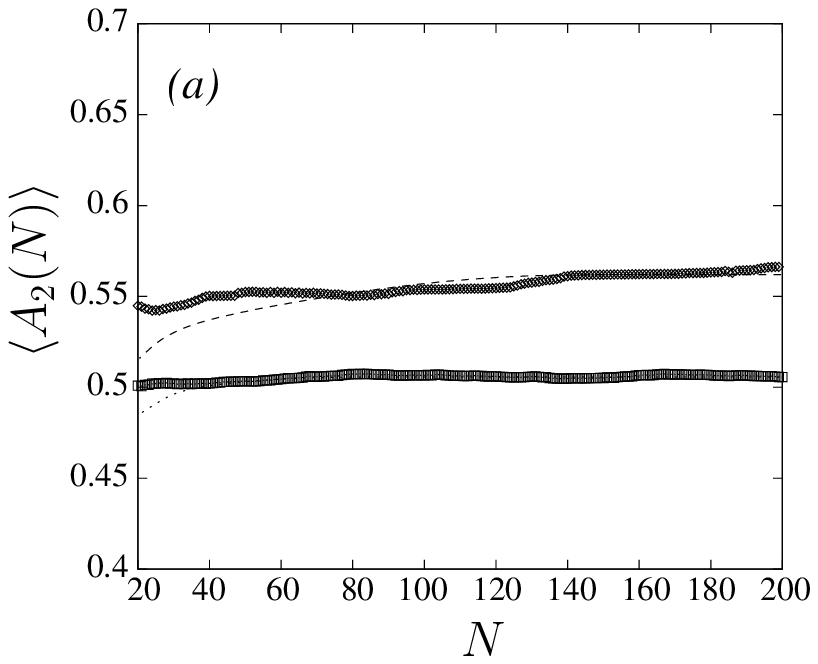}
\includegraphics[width=5.8cm]{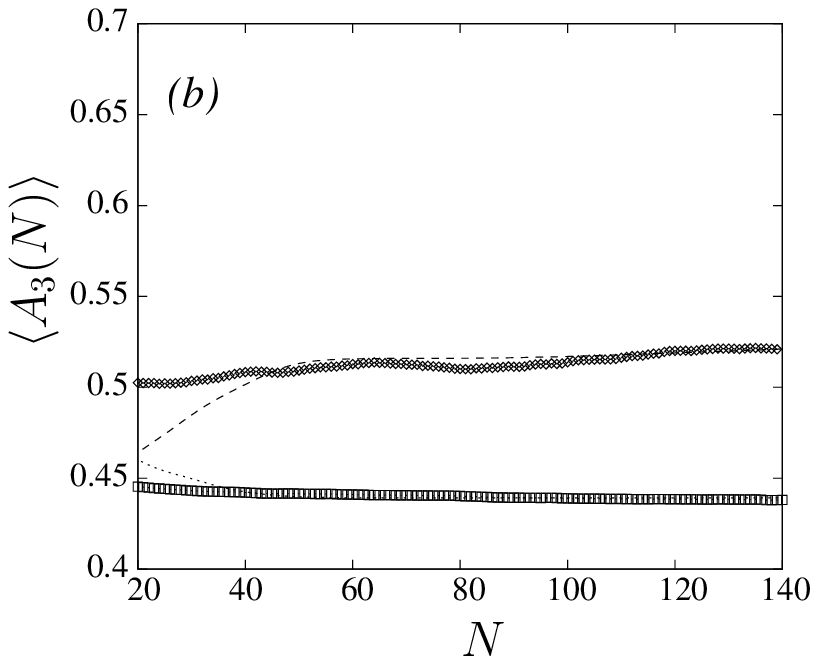}
\includegraphics[width=5.9cm]{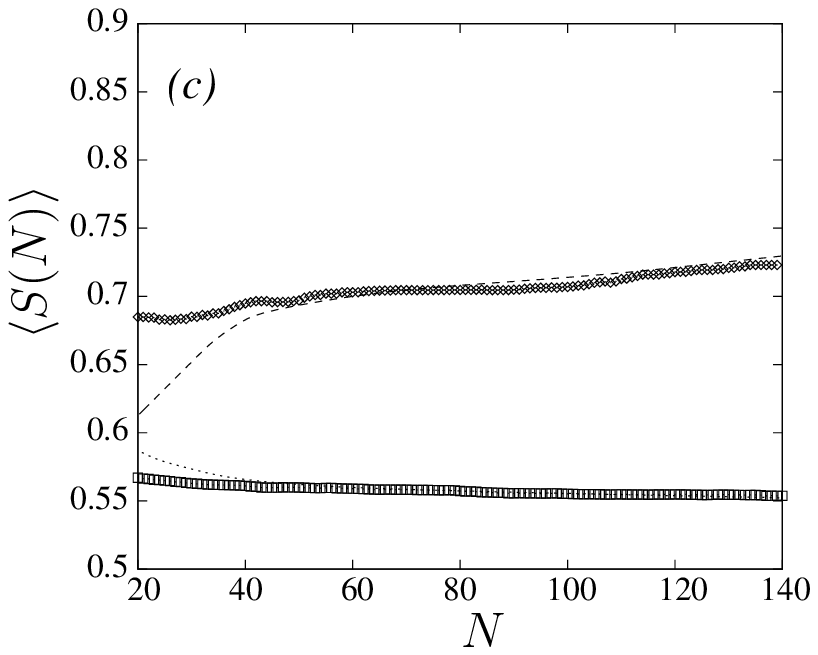}
\caption{\label{A}Averaged asphericity of polymer configurations in $d=2$ (a),  $d=3$ (b) and averaged prolateness in $d=3$ (c).
Lower lines: pure lattice, upper lines: percolation cluster. Dashed lines show the results of least-square fitting with
the ansatz (\ref{corr}).}
\end{figure}

\section{Results}

\begin{figure}[b!]
\includegraphics[width=5.8cm]{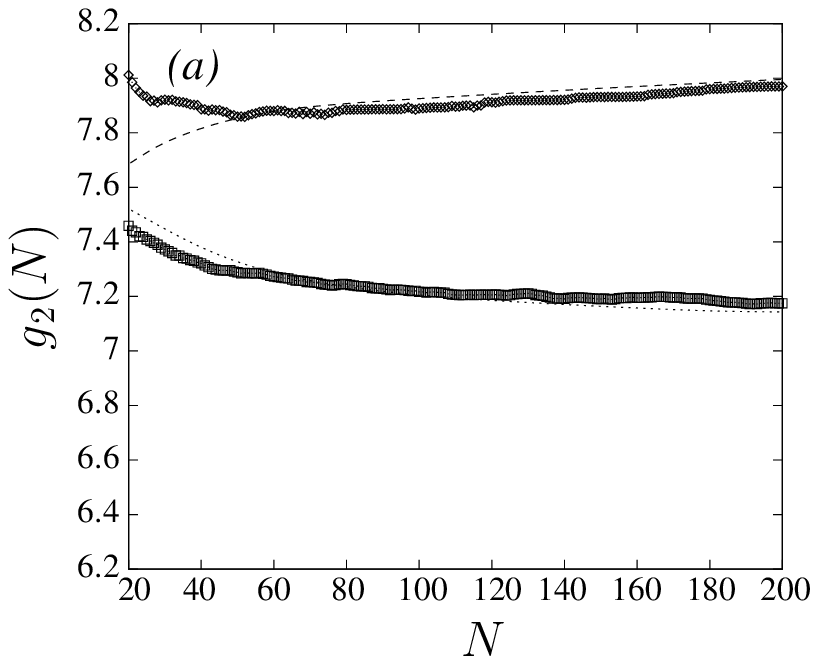}
\includegraphics[width=6.1cm]{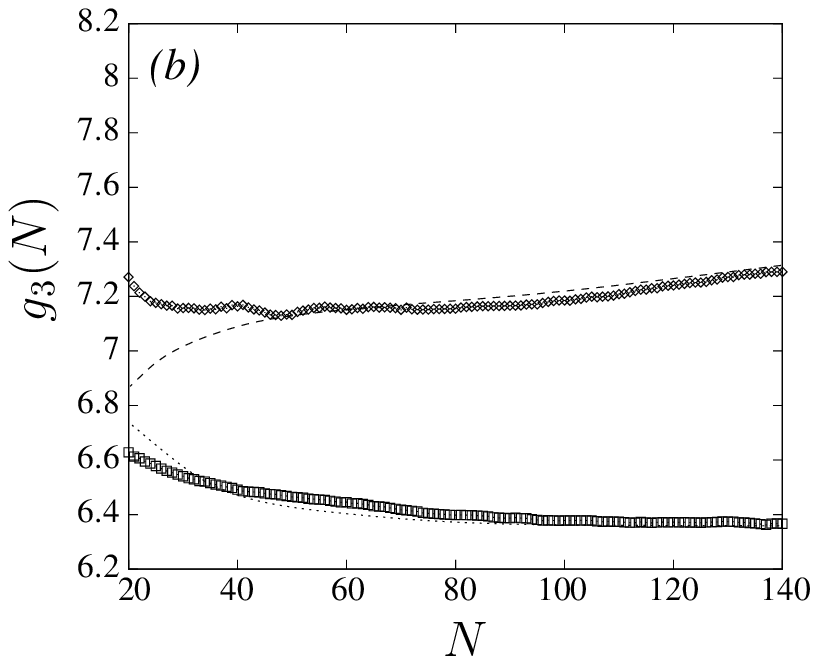}
\caption{\label{g}Size ratio $g_d\equiv\langle R_e^2 \rangle/\langle R_G^2 \rangle$ of polymer configurations in $d=2$ (a) and $d=3$ (b).
Lower lines: pure lattice, upper lines: percolation cluster. Dashed lines show the results of least-square fitting with
the ansatz (\ref{corr}).
  }
\end{figure}
We construct percolative lattices of edge lengths up to $L_{{\rm max}}{=}400,200$ in dimensions $d{=}2,3$, respectively, and
 estimate the mean shape parameters (\ref{avad}).
 The disorder averaging is performed over 1000 percolation clusters in each space dimension.

At first, let us analyze the connections (correlations) between values of the shape characteristics $A_d$ and $S$.
In Fig. \ref{correlation}(a) we present our data for simultaneous estimates of these quantities for each $N=120$-step SAW configuration of the ensemble in $d=3$.
As it is clear from the definitions (\ref{add}) and (\ref{sdd}), for spherical configurations both $A_3$ and $S_3$ are close to zero.
$S$ equals zero also for the cases when  $\lambda_1+\lambda_2=2\lambda_3$, corresponding to a non-zero asphericity value.
Increasing positive values of $S$ describe increasing the elongation and tending to rod-like structures,
which corresponds also to increasing  $A_3$ values. The negative values of $S$ describe the non-spherical oblate structures, also
corresponding to non-zero $A_3$ values. In Fig. \ref{correlation}(b)
we give also our data
for simultaneous estimates of  $R_G^2$ and $R_e^2$ in an ensemble of $N=120$-step SAWs in $d=3$; these quantities are less nicely correlated.

Figures \ref{A} and \ref{g} present simulation data for $\langle A_d \rangle$, $\langle S \rangle$ and $g_d\equiv\langle R_e^2 \rangle/\langle R_G^2 \rangle$
as functions of the chain
length $N$ in $d=2$, $3$. For comparison, we also evaluated the shape parameters on the pure lattices of the same edge lengths.
Note that the size ratio $g_d$ is rather a delicate quantity due to large fluctuations in the sample.

For the case of the pure lattice, it is evident that the shortest chains are very elongated in space and they become
a little more spherical with increasing chain length; thus $\langle A_d(N)\rangle$ and $\langle S(N)\rangle$ are decreasing functions of chain length $N$.
On a percolative lattice, these quantities behave in a different manner: they increase gradually with increasing $N$. The structure of the fractal percolation cluster
makes the longer polymer chain configurations to be more and more prolate.
The principal qualitative conclusions which we can derive from Figs. \ref{A} and \ref{g} is that
the shape parameters of typical polymer configurations change significantly relative to the obstacle-free case;
the shape tends to be more anisotropic, elongated due to the fractal structure of the lattice.

For finite chain length $N$, the values of shape parameters differ from those for infinitely long chains. This finite-size deviation obeys scaling behavior with $N$:
\begin{eqnarray}
\langle A_d(N) \rangle= \langle A_d \rangle +b_1N^{-\Delta}, \nonumber\\
\langle S(N) \rangle= \langle S \rangle +b_2N^{-\Delta}, \label{corr}\\
g_d(N)= g_d +b_3N^{-\Delta}, \nonumber
\end{eqnarray}
where $b_1$, $b_2$, $b_3$ are constants and $\Delta$ is the correction-to-scaling exponent: $\Delta(d=2)=1.5$ \cite{Caracciolo05},
$\Delta(d=3)=1.7$ \cite{Lam90}.
The shape parameter estimates can be obtained by least-square fitting of (\ref{corr}). For the case of the pure lattice, we receive a
nice agreement with
the existing data of Table \ref{dani}: $\langle A_2 \rangle=0.506\pm0.002$, $\langle A_3 \rangle = 0.435\pm0.002$,
 $\langle S \rangle = 0.545\pm0.002$.
Our results for the shape characteristics of SAWs on a percolation cluster are given in Table \ref{daniour}.
Note, that since  we can construct percolative lattice only up to a finite size $L$, it is not possible to perform very long SAWs on it.
For each $L$, the correct statistics  holds  only up to some ``marginal" number of SAWs steps
$N_{{\rm marg}}\sim L^{1/\nu_{{\rm SAW}}^{p_c}}$, with $\nu_{{\rm SAW}}^{p_c}$ being the size scaling exponent  on the percolation cluster
\cite {Blavatska08}.  We take this into account when analyzing the data obtained;
for each lattice size we are interested only in values of $N<N_{{\rm marg}}$, thus avoiding distortions, caused by finite-lattice effects.
We have also estimated the shape parameters defined by ratios of averages (\ref{avratio}) with $\langle \ldots \rangle$ replaced by $\overline{\langle \ldots\rangle}$ in the disordered case; as expected, the obtained values are considerably larger than
averages of ratios, cf. Table \ref{daniour}

\begin{figure}[t!]

\includegraphics[width=5.3cm]{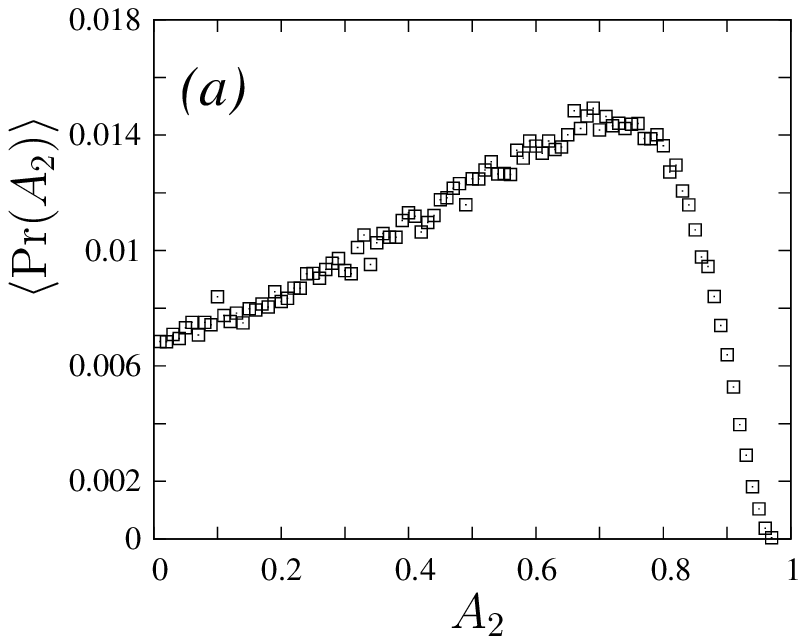}
\includegraphics[width=5.1cm]{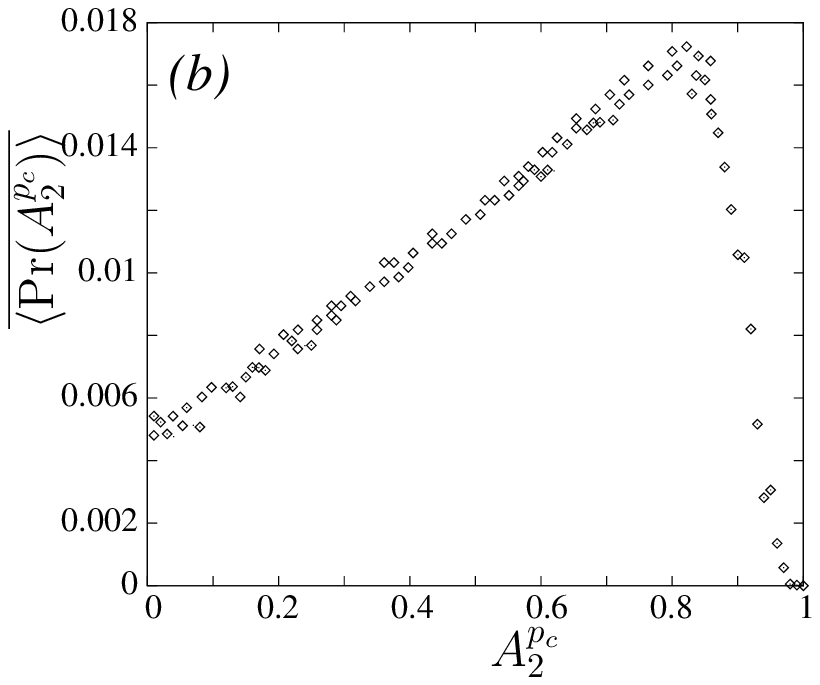}

\includegraphics[width=5.2cm]{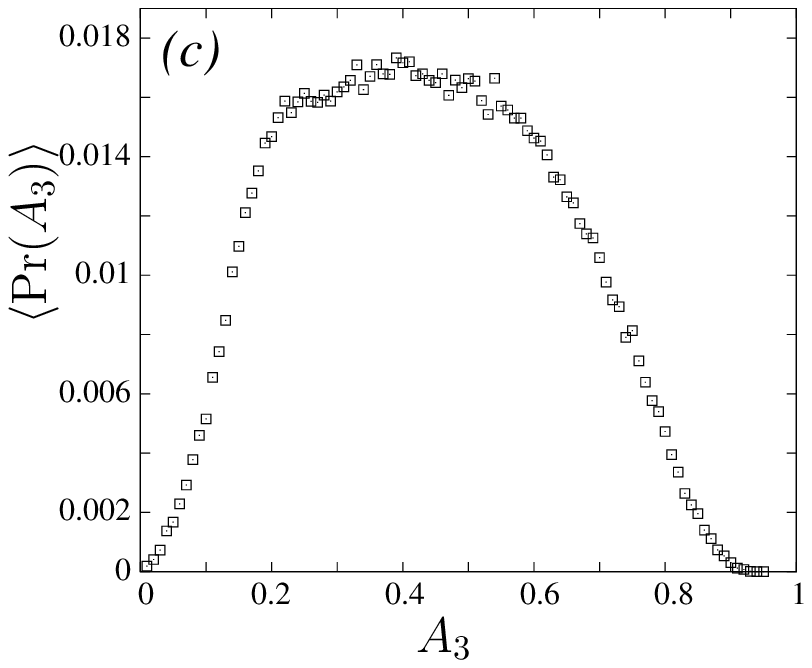}
\includegraphics[width=5.3cm]{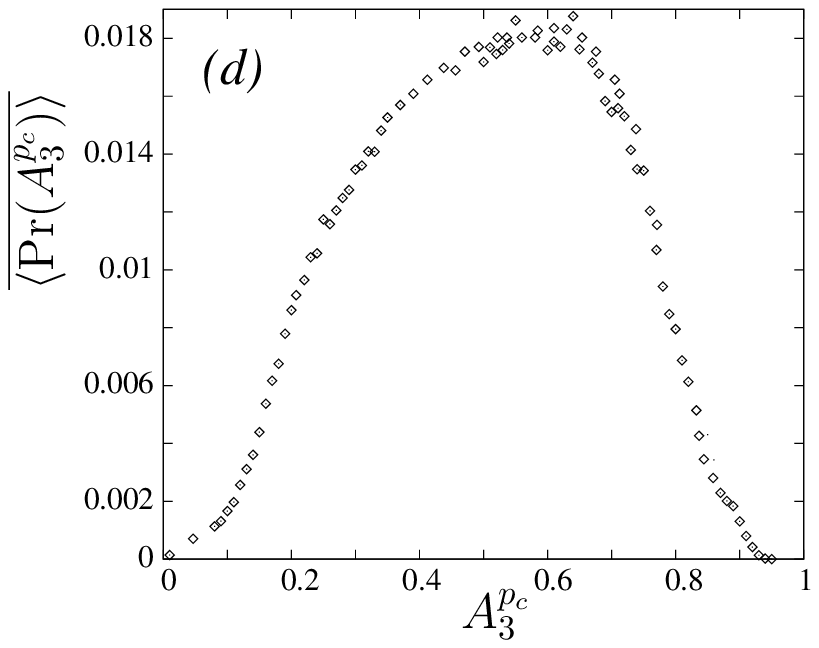}

\includegraphics[width=5.3cm]{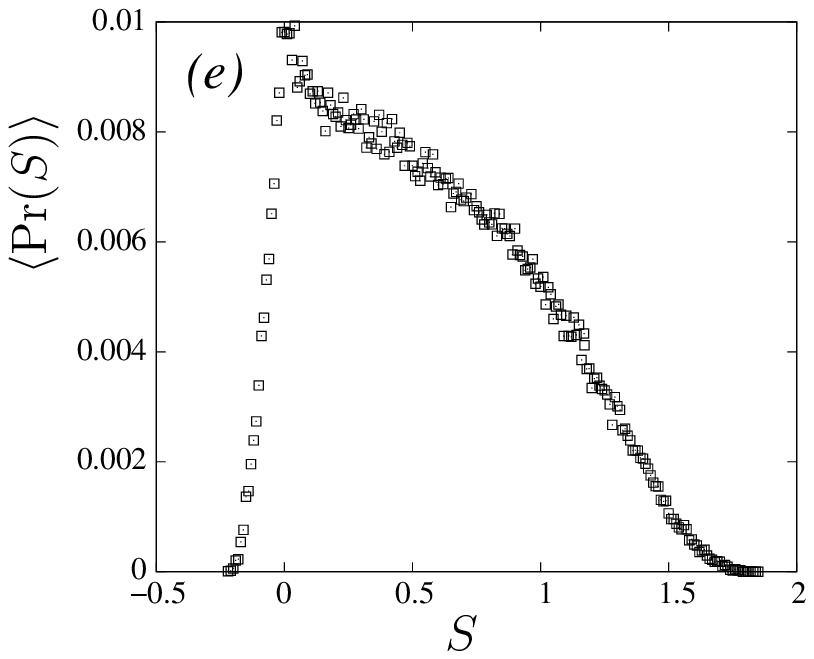}
\includegraphics[width=5.3cm]{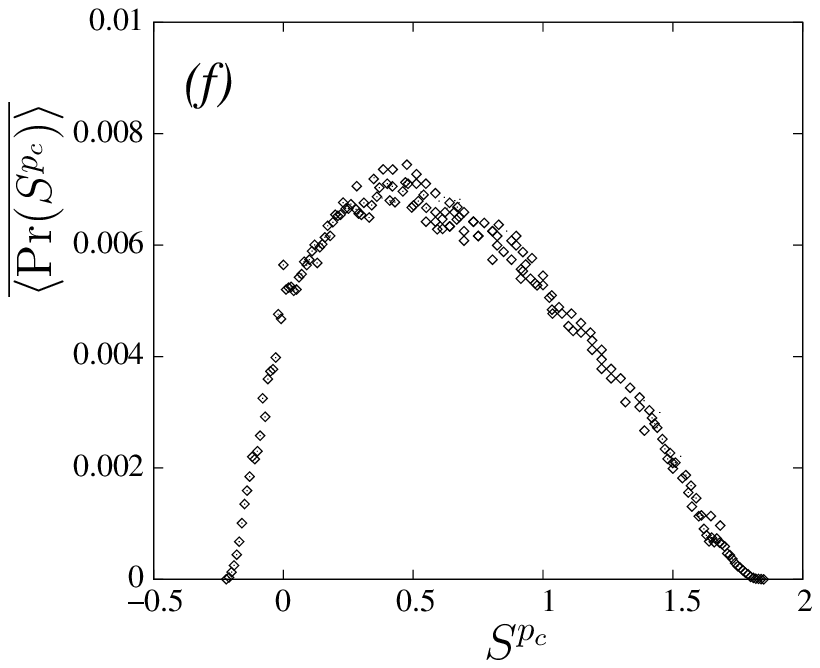}
\caption{\label{prob} Probability distributions of asphericity in $d=2$ ($a$, $b$) and $d=3$ ($c$, $d$) and prolateness in $d=3$
($e$, $f$) of SAWs
on a pure lattice (left column) and  on a percolation cluster (right column) for chain length $N=120$.
  }
\end{figure}
\begin{table}[t!]
\caption{ \label{daniour} Our estimates for size ratio, averaged asphericity  and prolateness of SAWs on percolation clusters.}
%\begin{center}
\begin{tabular}{cccccc}
\hline
 $d$ & $g_d^{p_c}$ & ${\overline {\langle A_d^{p_c}}} \rangle$& ${\overline{\langle S^{p_c}}} \rangle$ & $\hat{A}_d^{p_c}$ & $\hat{S}^{p_c}$\\
\hline
2 &  $7.96\pm0.01$ & $0.571\pm0.005$ &-- &  $0.68\pm0.01$ & --\\
3 &  $7.44\pm0.02$ & $0.531\pm0.005$ & $0.743\pm0.005$ & $0.61\pm0.01$ & $0.97\pm0.01$\\
\hline
\end{tabular}
%\end{center}
\end{table}

The probability distributions of shape parameters in $d=2,3$ at fixed chain length $N=120$ are given in Fig. \ref{prob}. The distribution functions for
$A_2$ and $A_2^{p_c}$ are rather unsymmetrical with the most probable value larger than the mean value. In $d=3$, the asphericity distribution function
is broad but quite symmetric, with the most probable and mean value being nearly equal.
The distribution function for $S$ is rather unsymmetrical with a most probable value of zero, which is shifted for $S^{p_c}$. The shapes
of these distributions indicate that the majority of polymer chain configurations has prolate asymmetry.

\section{Conclusions}

We study the universal size and shape characteristics of flexible polymer macromolecules in an environment with structural obstacles.
The measure of the shape properties of a specified configuration of the polymer chain is characterized in terms of the gyration tensor $\bf{Q}$. The rotationally invariant quantities, constructed as combinations of components of $\bf{Q}$,  such as the averaged asphericity $\langle A_d \rangle$ and prolateness $\langle S \rangle$ of typical chain realizations are of interest. Another quantity of interest is the universal size ratio $g_d\equiv \langle R_e^2\rangle / \langle R_G^2 \rangle$.
We address the question, how the polymer shape anisotropy is quantitatively influenced by the presence of disorder, which is important in understanding many real physical processes.

We use the lattice model of self-avoiding random walks on a disordered lattice exactly at the percolation threshold, when a percolation cluster with fractal structure emerges.  Studying processes on percolative lattices, one encounters two possibilities.
In the  first, one considers  only percolation clusters with linear size much larger than the typical length of the physical phenomenon under discussion (polymer chain length in
our case). The other statistical ensemble includes all lattice sites free of defects, which can be found in a percolative lattice.
We considered the former case, with a polymer chain residing only on the backbone of a percolation cluster.
Applying the Pruned-Enriched Rosenbluth Method, we performed computer simulations in $d=2$ and $d=3$ and obtained numerical estimates for the averaged asphericity, prolateness and
size ratio of self-avoiding walks on a percolation cluster. All the shape characteristics increase gradually with increasing polymer chain length; the structure of fractal percolation cluster
makes the longer polymer chain configurations to be more and more prolate.
Our results quantitatively indicate that the shape parameters of typical polymer configurations change significantly relative to the obstacle-free case; the shape tends to be more anisotropic and elongated due to the fractal structure of the lattice.

Note that recently a related model has been studied analytically in Ref.~\cite{Blavatska10}, where
the shape properties of polymers in a medium with obstacles, correlated at large separations $x$ according to a power law $x^{-a}$ were analyzed. Integer values of the parameter $a$
describe extended defects in the form of lines or planes of random orientation.
The obtained results qualitatively indicate an increase of shape asymmetry due to the presence of long-range correlated disorder, similarly to the case studied in the present paper.
In a forthcoming study, we are planning to confirm these results numerically and to obtain quantitative estimates of polymer characteristics in the presence of correlated extended defects.

\section{Acknowledgement}
V.B. is grateful for support through the S\"achsische DFG-Forschergruppe FOR877 and the  National Academy of Sciences of Ukraine Committee for young scientists.

\end{document}